\documentclass{PoS}

\usepackage{amsmath}

\newcommand{\be}{\begin{equation}}
\newcommand{\ee}{\end{equation}}
\newcommand{\bea}{\begin{eqnarray}}
\newcommand{\eea}{\end{eqnarray}}
\newcommand{\bt}{\begin{tabbing}}
\newcommand{\et}{\end{tabbing}}
\newcommand{\bi}{\begin{itemize}}
\newcommand{\ei}{\end{itemize}}
\newcommand{\ben}{\begin{enumerate}}
\newcommand{\een}{\end{enumerate}}

\newcommand{\calO}{{\mathcal O}}
\newcommand{\bfp}{{\bf p}}
\newcommand{\bfx}{{\bf x}}
\newcommand{\bfr}{{\bf r}}
\newcommand{\crad}{\langle r^2 \rangle}

\title{
   \begin{picture}(0,0)(0,0)%
   \put(355,75){\makebox(0,0)[l]{\textnormal{\normalsize KEK-CP-197}}}%
   \end{picture}%
   Pion form factor from all-to-all propagators of overlap quarks
}

\ShortTitle{
   Pion form factor from all-to-all propagators of overlap quarks
}

\author{ 
   JLQCD collaboration: 
   \speaker{T.~Kaneko}$^{a,b}$\thanks{E-mail: takashi.kaneko@kek.jp}, 
   H.~Fukaya$^c$, 
   S.~Hashimoto$^{a,b}$, 
   H.~Matsufuru$^{a}$, 
   J.~Noaki$^{a}$, 
   T.~Onogi$^{d}$
   and
   N.~Yamada$^{a,b}$
   \\
   \\
   \llap{$^a$}
   High Energy Accelerator Research Organization (KEK),
   Ibaraki 305-0801, Japan 
   \\
   \llap{$^b$}
   School of High Energy Accelerator Science,
   The Graduate University for Advanced Studies (Sokendai),
   Ibaraki 305-0801, Japan
   \\
   \llap{$^c$}
   Theoretical Physics Laboratory, RIKEN, Saitama 351-0198, Japan
   \\
   \llap{$^d$}
   Yukawa Institute for Theoretical Physics, Kyoto University,
   Kyoto 606-8502, Japan
}

\abstract{
We report on our calculation of the pion electromagnetic form factor 
with two-flavors of dynamical overlap quarks. 
Gauge configurations are generated using the 
Iwasaki gauge action on a $16^3 \times 32$ lattice at the lattice spacing 
of 0.12~fm with sea quark masses down to $m_s/6$, where
$m_s$ is the physical strange quark mass. 
We describe our setup to measure the form factor through 
all-to-all quark propagators and present preliminary results.
}

\FullConference{
   The XXV International Symposium on Lattice Field Theory\\
   July 30-4 August 2007\\
   Regensburg, Germany
}

\begin{document}


\section{Introduction} 

%


Since the pion plays a central role in low-energy dynamics, 
understanding its properties is of great importance.
For the electromagnetic form factor $F_\pi(q^2)$, 
precise experimental data are available near 
the zero momentum transfer $q^2\!=\!0$, 
where 
the dependence of $F_\pi(q^2)$ on the quark mass $m$ and $q^2$ 
can be described by chiral perturbation theory (ChPT)  
provided that $m$ is sufficiently small.
A detailed comparison of $F_\pi(q^2)$ on the lattice
with ChPT and experiments therefore provides 
a good testing ground for 
lattice calculations in the chiral regime.
An understanding on the applicability of ChPT to lattice data
is also helpful towards a reliable calculation of form factors 
of $K$, $D$ and $B$ mesons.


In this article,
we report on our on-going calculation of $F_\pi(q^2)$ in two-flavor QCD.
We employ the overlap fermions, 
which have the exact chiral symmetry and hence 
allow us to apply ChPT straightforwardly to our chiral extrapolation.
The salient feature of this study is that 
$F_\pi(q^2)$ is calculated precisely through all-to-all quark propagators
\cite{A2A} for a meaningful comparison with ChPT and experiments.


\section{Simulation method}


We simulate QCD with two flavors of degenerate up and down quarks
using the Iwasaki gauge action and the overlap quark action 
with the standard Wilson Dirac kernel.
To reduce the computational cost substantially,
(near-)zero modes of the kernel are suppressed by introducing 
two-flavors of unphysical Wilson fermions and twisted mass
ghosts \cite{exW+extmW:JLQCD},
which do not change the continuum limit.
Our numerical simulations are carried out 
on a $N_s^3 \times N_t \!=\! 16^3 \times 32$ lattice at a single value of $\beta\!=\!2.30$.
The lattice spacing is $a\!=\!0.1184(16)$~fm,
if $r_0\!=\!0.49$~fm is used as input.
We take four quark masses $m\!=\!0.015,0.025,0.035$ and 0.050, 
which cover a range of $[m_s/6,m_s/2]$.
%
Our current statistics are 50 configurations 
separated by 100 HMC trajectories at each $m$.
So far, we have simulated only the trivial topological sector,
and effects of the fixed global topology by the extra Wilson fermions
are to be studied \cite{fixedQ}.
We refer to Ref.\cite{Lat07:JLQCD:Matsufuru}
for further details on our production run.


\section{Measurement through all-to-all propagators}


We construct all-to-all propagators of overlap quarks 
along the strategy proposed in Ref.~\cite{A2A}.
Low-lying modes of the overlap operator $D$ are determined 
by the implicitly restarted Lanczos algorithm
and their contribution to the all-to-all propagator 
is calculated exactly as 
\bea
   (D^{-1})_{\rm low}
   & = &
   \sum_{k=1}^{N_{\rm ep}}
      \frac{1}{\lambda^{(k)}}\,u^{(k)}u^{(k)\dagger},
   \label{eqn:a2a_prop:low}
\eea
where $(\lambda^{(k)},u^{(k)})$ represents $k$-th eigenmode
and the number of eigenmodes $N_{\rm ep}$ is set to 100 in this study.
We note that the overlap operator is normal and 
the left and right eigenvectors coincide with each other.



The contribution of higher modes is estimated stochastically 
by the noise method with the dilution technique \cite{A2A}.
One $Z_2$ noise vector is generated for each configuration,
and is {\it diluted} into
\noindent
$N_d = 3 \times 4 \times N_t/2$ 
vectors with support on a single value for color and spinor 
indices and at two consecutive time-slices.
The high mode contribution 
\bea
   (D^{-1})_{\rm high} 
   & = & 
   \sum_{d=1}^{N_d} x^{(d)}\,\eta^{(d) \dagger}
   \label{eqn:a2a_prop:high}
\eea
can be obtained by solving the linear equation for each diluted source
\bea
   D\,x^{(d)} 
   = (1-P_{\rm low}) \, \eta^{(d)} 
   \hspace{1mm}           
   (d=1,...,N_d),
   \hspace{2mm}           
   \label{eqn:a2a_prop:high:leq}
\eea
where $d$ is the index for the dilution and 
$P_{\rm low}$ is the projector to the eigenspace spanned by
the low modes. We employ the four dimensional relaxed CG
for our overlap solver \cite{relCG}.

In summary, 
all-to-all quark propagators can be expressed as the matrix
\bea
   D^{-1}
   & = &
   \sum_{k=1}^{N_{\rm vec}} v^{(k)}\,w^{(k)\dagger}
   \hspace{5mm} (N_{\rm vec}=N_{\rm ep}+N_d)
   \label{eqn:a2a_prop}
\eea
constructed from the following two set of vectors $v$ and $w$:
\bea
   v^{(k)} 
   & = & 
   \left\{
      \frac{u^{(1)}}{\lambda^{(1)}},
      \ldots,
      \frac{u^{(N_{\rm ep})}}{\lambda^{(N_{\rm ep})}},
      x^{(1)}, \dots, x^{(N_d)}
   \right\},
   \hspace{3mm}
   w^{(k)} 
   = 
   \left\{
      u^{(1)}, \ldots, u^{(N_{\rm ep})},
      \eta^{(1)}, \dots, \eta^{(N_d)}
   \right\}.
   \label{eqn:vw_vectors}
\eea



Then,
two-point functions 
with the source (sink) operator at time-slice $t^{(\prime)}$ 
and three-point functions with the vector current at $t^{\prime\prime}$ 
can be expressed as 
\bea
    C_{\Gamma \Gamma^\prime, \phi \phi^\prime}(t^\prime-t;\bfp)
    & =  &
    \sum_{k,l=1}^{N_{\rm vec}}
    \calO^{(k,l)}_{\Gamma^\prime,\phi^\prime}(t^\prime,\bfp) \, 
    \calO^{(l,k)}_{\Gamma,\phi}(t,-\bfp),
    \label{eqn:msn_corr_2pt} 
    \\
    C_{\Gamma \gamma_\mu \Gamma^\prime, \phi \phi^\prime}
    (t^{\prime\prime}-t,t^\prime-t^{\prime\prime};\bfp,\bfp^\prime)
    & = & 
    \sum_{k,l,m=1}^{N_{\rm vec}}
    \calO^{(m,l)}_{\Gamma^\prime,\phi^\prime}(t^\prime,\bfp^\prime) \, 
    \calO^{(l,k)}_{\gamma_\mu,\phi_{l}}(t^{\prime\prime},\bfp-\bfp^\prime) \, 
    \calO^{(k,m)}_{\Gamma,\phi}(t,-\bfp),
    \label{eqn:msn_corr_3pt}  
\eea
where the momentum and smearing function for the initial (final) meson
are denoted by $\bfp^{(\prime)}$ and $\phi^{(\prime)}$, 
and 
\bea
   \calO^{(k,l)}_{\Gamma,\phi}(t,\bfp)
   & = & 
   \sum_{\bfx,\bfr}
   \phi(\bfr)\, 
   w(\bfx+\bfr,t)^{(k)\dagger} \, 
   \Gamma \,
   v(\bfx,t)^{(l)}\,
   e^{-i \bfp \bfx}
   \label{eqn:meson_op}
\eea
is the meson operator with the Dirac spinor structure $\Gamma$ 
constructed from the $v$ and $w$ vectors.
The smearing function for the local operator
is $\phi_l(\bfr)\!=\!\delta_{{\bf r},{\bf 0}}$.


We prepare the $v$ and $w$ vectors on the IBM BlueGene/L at KEK.
The computational cost of the determination of low modes
is $\sim 0.6~\mbox{TFLOPS} \cdot \mbox{hours}$ 
per configuration.
Solving Eq.~(\ref{eqn:a2a_prop:high:leq}) is the most time-consuming part 
in our measurement, 
since it requires $N_t/2$ times more inversions than the conventional method.
We observe that, however, 
the low-mode preconditioning of our overlap solver
leads to about a factor of 8 speedup 
and its cost is reduced to 
$\sim 1.7~\mbox{TFLOPS} \cdot \mbox{hours} / \mbox{conf}$
for a given valence quark mass $m$.
The calculation of the meson operator $\calO^{(k,l)}_{\Gamma,\phi}(t,\bfp)$
needs much less CPU time than the above two steps:
it is about $0.2~\mbox{GFLOPS} \cdot \mbox{hours} / \mbox{conf}$
for a single choice of ($m$,$\bfp$,$\Gamma$,$\phi$).
The calculation of correlation functions
%
are even less costly.
These calculations are carried out on the Hitachi SR11000 and workstations
at KEK.


\FIGURE{
   \centering
   \includegraphics[angle=0,width=0.45\linewidth,clip]{jkd_msn_corr_3pt.eps}
   \vspace{-2mm}
   \caption{
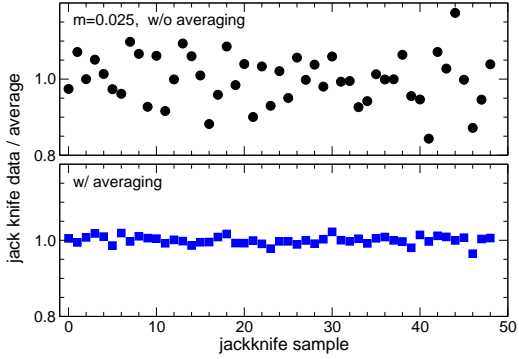
      Jackknife data of three-point function 
      $C_{\gamma_5 \gamma_4 \gamma_5, \phi_s \phi_s}(N_t/4,N_t/4;\bfp,{\bf 0})$
      with $|\bfp|\!=\!\sqrt{2}$
      before (top panel) and after averaging over source operator locations
      and momentum configurations (bottom panel). 
      Data are normalized by the statistical average.
   }
   \label{fig:a2a:3pt:jkd}     
}

The key issue in the all-to-all calculation is the re-usability 
of the all-to-all propagators:
namely,
we do not have to repeat the time-con\-sum\-ing Lanczos step and overlap
solver to construct the meson operator
$\calO^{(k,l)}_{\Gamma,\phi}(t,\bfp)$
for different choices of ($\bfp$,$\Gamma$,$\phi$).
This is a great advantage in studies of form factors,
which require an accurate estimate of relevant correlation functions
with various choices of the momentum configuration $(\bfp,\bfp^\prime)$.
In this study, 
we test two smearing functions $\phi_{l}(\bfr)$ and 
$\phi_{s}(\bfr)\!=\!\exp[-0.4|\bfr|]$,
and take 33 choices for the meson momentum $\bfp$ 
with $|\bfp|\!\leq\!2$.
Note that the lattice momentum is in units of $2\pi/L$ in this article.
This setup enables us to simulate 11 different values of $q^2$,
which cover a range of $-1.65~\mbox{[GeV$^2$]} \lesssim q^2$.


It is also advantageous to 
average the correlation functions over the momentum 
configurations, which give the same value of $q^2$, 
as well as over the source locations $({\bf x},t)$
with temporal separations,
namely $\Delta t\!=\!t^{\prime\prime}\!-\!t$ 
and $\Delta t^\prime\!=\!t^\prime\!-\!t^{\prime\prime}$,
kept fixed.
This averaging reduces the statistical fluctuation 
remarkably
as shown in Fig.~\ref{fig:a2a:3pt:jkd}.


\section{Pion form factor and charge radius}


\begin{figure}[b]
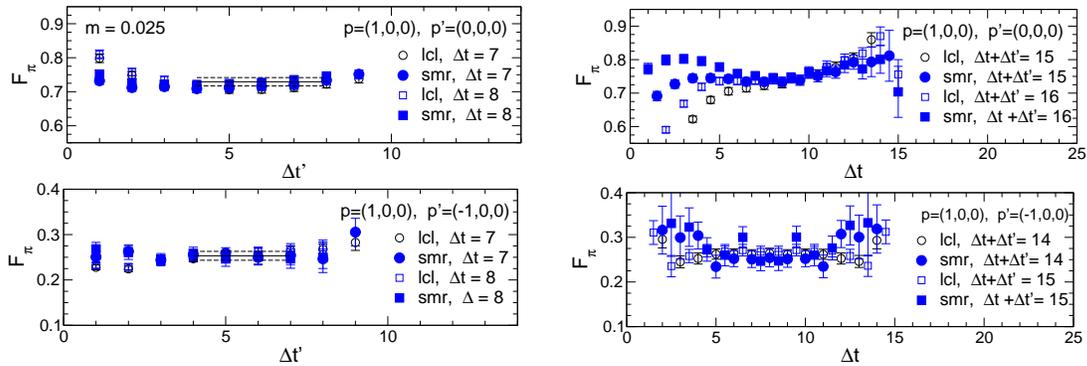

\begin{center}
\includegraphics[angle=0,width=0.45\linewidth,clip]%
                {pff_vs_dtsnk.mom0100.eps}
\hspace{5mm}
\includegraphics[angle=0,width=0.45\linewidth,clip]%
                {pff_vs_dtsrc.mom0100.eps}

\includegraphics[angle=0,width=0.45\linewidth,clip]%
                {pff_vs_dtsnk.mom0101.eps}
\hspace{5mm}
\includegraphics[angle=0,width=0.45\linewidth,clip]%
                {pff_vs_dtsrc.mom0101.eps}
\vspace{-2mm}
\caption{
   Effective value of pion form factor 
   $F_{\pi,\phi}(\Delta t,\Delta t^\prime;q^2)$
   at $m\!=\!0.025$.
   In the left panels,
   the data are plotted as a function of $\Delta t^\prime$ 
   with $\Delta t$ fixed,
   whereas the right panels show $\Delta t$ dependence with 
   $\Delta t + \Delta t^\prime$ fixed.
   Open (filled) symbols show data with $\phi\!=\!\phi_{l}$ 
   ($\phi_s$).
}
\label{fig:pff:pff_eff}
\end{center}
\end{figure}


We calculate effective value of the pion form factor from the ratio 
\bea
   F_{\pi,\phi}(\Delta t,\Delta t^\prime;q^2)
   & = & 
   \frac{2\,M_\pi}{E_\pi(|\bfp|)+E_\pi(|\bfp^\prime|)}
   \frac{R_{\phi}(\Delta t,\Delta t^\prime; |\bfp|,|\bfp^\prime|,q^2)}
        {R_{\phi}(\Delta t,\Delta t^\prime; 0,0,0)},
   \\
   R_\phi(\Delta t,\Delta t^\prime; |\bfp|, |\bfp^\prime|, q^2)
   & = &
   \frac{C_{\gamma_5\gamma_4\gamma_5,\phi \phi}
         (\Delta t,\Delta t^\prime; \bfp,\bfp^\prime)}
        {C_{\gamma_5\gamma_5,\phi \phi_l}(\Delta t;\bfp)\,
         C_{\gamma_5\gamma_5,\phi_l \phi}(\Delta t^\prime;\bfp^\prime)},
   \label{eqn:pff:ratio}
\eea       
where 
$\phi = \phi_{l}$ or $\phi_{s}$,
and the pion mass $M_\pi$ and energy $E_\pi$ are determined 
by single-cosh fits to 
$C_{\gamma_5\gamma_5,\phi_{s} \phi_{s}}$.
We note that the ratio $R_\phi$ is calculated from correlation functions 
averaged over the momentum configurations and source locations.

\begin{figure}[t]
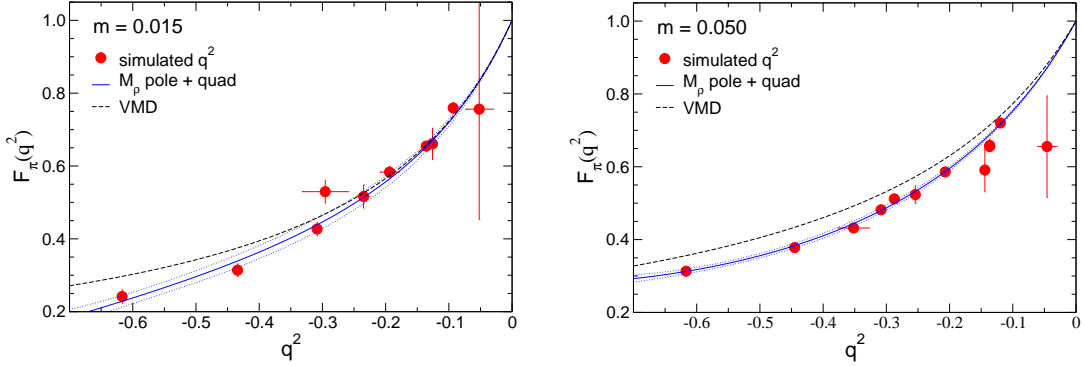

\begin{center}
\includegraphics[angle=0,width=0.45\linewidth,clip]%
                {pff_vs_q2.m0015.eps}
\hspace{5mm}
\includegraphics[angle=0,width=0.45\linewidth,clip]%
                {pff_vs_q2.m0050.eps}
\vspace{-2mm}
\caption{
   Pion form factor at $m\!=\!0.015$ (left panel) and 0.050 (right panel)
   as a function of $q^2$.
   The Solid line shows the parametrization 
   of the measured pole plus the quadratic correction.
   The dashed line is the expectation from VMD.
}
\label{fig:pff:pff_vs_q2}
\vspace{-6mm}
\end{center}
\end{figure}

An example of $F_{\pi,\phi}(\Delta t,\Delta t^\prime;q^2)$
is plotted in Fig.~\ref{fig:pff:pff_eff}.
The pion form factor $F_\pi(q^2)$ is determined from a constant 
fit to $F_{\pi,\phi_{s}}(\Delta t,\Delta t^\prime;q^2)$ 
in a range of $(\Delta t,\Delta t^\prime)$, 
where $F_{\pi,\phi_{s}}$ 
shows a reasonable plateau and 
good agreement with data with $\phi\!=\!\phi_l$.
As shown in Fig.~\ref{fig:pff:pff_vs_q2},
we obtain an accurate estimate of $F_\pi(q^2)$ except at our smallest $q^2$, 
where $C_{\gamma_5\gamma_4\gamma_5,\phi \phi}$
suffers from the most serious damping factor
$e^{-E_\pi(|\bfp|)\Delta t}e^{-E_\pi(|\bfp^{\prime}|)\Delta t^{\prime}}$
with $(|\bfp|,|\bfp^{\prime}|)\!=\!(2,1)$.

In the same figure,
we observe that the $q^2$ dependence of our data is close to the expectation 
from the vector meson dominance (VMD) hypothesis
$F_\pi(q^2)\! \sim \! 1/(1-q^2/M_{\rho}^2)$
particularly near $q^2\!=\!0$.
%
%
The $q^2$ dependence is therefore parametrized 
by the following form of the vector meson pole
with a polynomial (up to cubic order) 
\bea
   F_\pi(q^2)
   & = & 
   \frac{1}{1-q^2/M_{\rho}^2} + c_1\,q^2 + c_2\,q^4 + c_3\,q^6,
   \label{eqn:pff:q2_fit:poly}
\eea
or an additional pole correction
\bea
   F_\pi(q^2)
   & = & 
   \frac{c}{1-q^2/M_{\rho}^2} + \frac{c^\prime}{1-q^2/M_{\rm pole}^2}
   \hspace{5mm}
   (c+c^\prime\!=\!1).
   \label{eqn:pff:q2_fit:2poles}
\eea
While the simplest form Eq.~(\ref{eqn:pff:q2_fit:poly})
with the linear correction ($c_2,c_3\!=\!0$)
gives a slightly high value of $\chi^2/{\rm dof} \gtrsim 2$
at heavier quark masses $m\!\geq\!0.035$,
other fitting forms describe our data reasonably well at all $m$.


\FIGURE{
   \centering
   \includegraphics[angle=0,width=0.25\linewidth,clip]{ifit_vs_r2.eps}
   \includegraphics[angle=0,width=0.22\linewidth,clip]{q2min_vs_r2.eps}
   \vspace{-2mm}
   \caption{
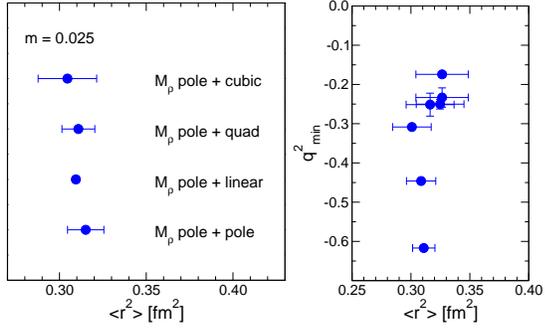
      Charge radius obtained at $m\!=\!0.025$ 
      from various choices of fitting form (left panel)  
      and lower cut for fit range
      in parametrization of $q^2$ dependence of $F_\pi(q^2)$  
      (right panel).
      \vspace{5mm}
   }
   \label{fig:pff:r2}
}

In Fig.~\ref{fig:pff:r2}, we compare the charge radius
\bea
   \crad
   & = & 
   \left. 6\,dF_\pi(q^2)/dq^2 \right|_{q^2=0}
   \label{eqn:pff:r2:def}
\eea
obtained from different fitting forms and ranges 
for the parametrization of the $q^2$ dependence.
Our result is quite stable against variation of these fitting setup.
In the following, 
we employ Eq.~(\ref{eqn:pff:q2_fit:poly}) with the quadratic correction,
since it gives the least value of $\chi^2/{\rm dof}$ with reasonably 
well-determined fitting parameters.
We include the leading finite volume correction \cite{FSE}
into the result for $\crad$.

\begin{figure}[b]
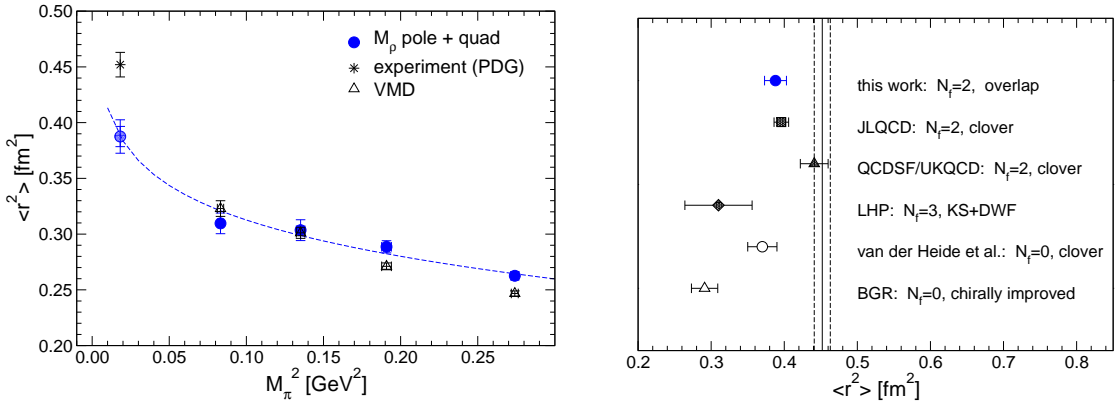

\begin{center}
\includegraphics[angle=0,width=0.485\linewidth,clip]{r2_vs_Mpi2.eps}
\hspace{5mm}
\includegraphics[angle=0,width=0.445\linewidth,clip]{r2_compare.eps}

\vspace{-2mm}
\caption{
   Left panel: 
   chiral extrapolation of charge radius $\crad$.
   The experimental value in Ref.\cite{r2:exprt} 
   and  $\crad\!=\!6/M_\rho^2$ from VMD are alto plotted.
   Right panel: 
   comparison of $\crad$ from recent studies.
}
\label{fig:pff:r2:chiral_fit}
\end{center}
\end{figure}

Figure~\ref{fig:pff:r2:chiral_fit} shows 
our chiral extrapolation of $\crad$.
In this preliminary report, 
we test the NLO ChPT formula \cite{PFF:ChPT:NLO} 
\bea
   \crad
   = 
   c_0  
   + 
   \frac{1}{(4 \pi\,f_0)^2} \log \left[ M_\pi^2 \right]
   +
   c_1\,M_\pi^2,
   \hspace{5mm}
   \label{eqn:pff:chiral_fit}
\eea
where 
a higher order analytic correction is included
to account for the quark mass dependence of the contribution of 
the vector resonance $6/M_\rho^2$.
With two values of $f_0$ from our studies in $p$- and $\epsilon$-regimes
\cite{Lat07:JLQCD:Noaki-Fukaya}, 
Eq.~(\ref{eqn:pff:chiral_fit}) gives a reasonable value of 
$\chi^2/{\rm dof}\!\sim\!1.2$
even without the higher order term.
It is however likely that this consistency with NLO ChPT is accidental,
since, as seen Fig.~\ref{fig:pff:r2:chiral_fit}, 
the quark mass dependence of our data is mainly caused by 
that of the resonance contribution.

This chiral extrapolation 
leads to our preliminary result
\bea
   \crad
   & = &
   0.388(9)_{\rm stat}(12)_{\rm sys}~\mbox{fm}^2,
   \label{eqn:pff:r2:rst}
\eea
where the systematic error is estimated by 
changing the parametrization form of 
the $q^2$ dependence of $F_\pi(q^2)$
and the choice of $f_0$,
and by removing the higher order correction
in Eq.~(\ref{eqn:pff:chiral_fit}).
This result is significantly smaller than the experimental value
0.452(11)~fm \cite{r2:exprt}.
We need further investigations on systematic uncertainties:
namely the lattice scale has to be fixed from an experimental input
and we need study finite volume effects 
including those due to the fixed topology \cite{fixedQ}.
The consistency with ChPT may also be tested 
within the framework including resonance contributions
\cite{rChPT}
as in an analysis of experimental data in Ref~\cite{PFF:ChPT:NNLO}.


\section{Conclusions} 


In this article,
we report on our calculation of $F_\pi(q^2)$
in two-flavor QCD 
through all-to-all propagators of the overlap fermions.
Our preliminary result for $\crad$ is significantly smaller
than experiment
as in most of previous studies 
\cite{PFF:Nf2:Plq+Clv:HKL,PFF:impG+AT+DWF:LHP,PFF:Nf2:Plq+Clv:JLQCD,PFF:Nf0:LW+CImpBGR,PFF:Nf2:Plq+Clv:QCDSF,PFF:review}
shown in Fig.~\ref{fig:pff:r2:chiral_fit}.
To understand the source of this discrepancy, 
we are completing our measurement of $F_\pi(q^2)$ 
with our full statistics (10,000 trajectories at each $m$)
for a more stringent comparison with experiment and ChPT.

We also observe that the all-to-all propagators 
provide a very precise determination of meson correlators.
Our studies are already underway for the pion scalar form factor, 
$K \to \pi$ form factors and flavor singlet mesons 
using meson operators Eq.~(\ref{eqn:meson_op}) stored on the disk.


\vspace{5mm}

Numerical simulations are performed on Hitachi SR11000 and 
IBM System Blue Gene Solution 
at High Energy Accelerator Research Organization (KEK) 
under a support of its Large Scale Simulation Program (No.~07-16).
This work is supported in part by the Grant-in-Aid of the
Ministry of Education 
(No.~17740171, 18034011, 18340075, 18740167, 18840045, 19540286 and 19740160).


\end{document}